\documentclass[a4paper,fleqn]{cas-sc}

\usepackage[authoryear]{natbib}
\usepackage{xcolor}
\usepackage{longtable}
\usepackage{booktabs}
\usepackage{array}

\begin{document}
\let\WriteBookmarks\relax
\def\floatpagepagefraction{1}
\def\textpagefraction{.001}

\shorttitle{Design of Nuclear and Magnetic SLD Profiles}

\shortauthors{Barnasas et~al.}

\title[mode=title]{Design of Nuclear and Magnetic Scattering Length Density Profiles}

\author[1]{Alexandros Barnasas}[orcid=0000-0002-5523-0714]
\credit{Conceptualization, Data Curation, Investigation, Methodology, Resources, Validation, Writing -- Original Draft}

\affiliation[1]{organization=Department of Physics and Astronomy, Uppsala University,
            addressline={Box 516},
            city={Uppsala},
            postcode={SE-75120},
            country={Sweden}}

\author[1]{Edward J. Stuckey}[orcid=0000-0003-1489-4079]
\cormark[1]
\ead{Edward.Stuckey@physics.uu.se}
\credit{Data Curation, Formal analysis, Methodology, Software, Visualization, Writing -- Original Draft}

\author[1,2]{Alexei A. Vorobiev}[orcid=0000-0003-3778-4948
]
\credit{Data Curation, Investigation, Validation, Writing -- Review \& Editing}

\affiliation[2]{organization={Institut Laue Langevin},
            addressline={Avenue des Martyrs 71},
            city={Grenoble},
            postcode={38000},
            country={France}}

\author[3]{Joshaniel F.K. Cooper}[orcid=0000-0002-4635-2381]
\credit{Formal analysis, Writing -- Review \& Editing}

\affiliation[3]{organization={European Spallation Source ERIC},
            city={Lund},
            country={Sweden}}

\author[1]{Sjoerd Stendahl}[orcid=0000-0002-6571-0746]
\credit{Investigation, Writing -- Review \& Editing}

\author[1]{Bj\"orgvin Hj\"orvarsson}[orcid=0000-0003-1803-9467]
\credit{Conceptualization, Funding acquisition, Methodology, Project administration, Resources, Supervision, Writing -- Original Draft}

\cortext[1]{Corresponding author}
\begin{abstract}
The scattering length density (SLD) profile of a sample can be deliberately designed to enhance the information of interest obtained from neutron reflectometry. Here, we demonstrate this approach using amorphous $\mathrm{Fe}_x\mathrm{Co}_y\mathrm{Al}_8\mathrm{Zr}_2$ thin films. Four samples, each comprising ten repetitions of an AlZr spacer layer and a magnetic layer, were prepared with magnetic-layer compositions of $\mathrm{Fe}_{80}\mathrm{Co}_{10}\mathrm{Al}_8\mathrm{Zr}_2$, $\mathrm{Fe}_{70}\mathrm{Co}_{20}\mathrm{Al}_8\mathrm{Zr}_2$, $\mathrm{Fe}_{60}\mathrm{Co}_{30}\mathrm{Al}_8\mathrm{Zr}_2$, and $\mathrm{Fe}_{45}\mathrm{Co}_{45}\mathrm{Al}_8\mathrm{Zr}_2$, and characterised using X-ray and polarised neutron reflectometry. Across the series, the nuclear SLD of the FeCoAlZr alloy decreases approximately linearly from $7.14\times10^{-6}$ to $4.63\times10^{-6}$~\AA$^{-2}$ with increasing Co content, while the magnetic SLD is approximately $4.3$--$4.5\times10^{-6}$~\AA$^{-2}$. These results demonstrate that the nuclear and magnetic contributions can be controlled largely independently, with composition primarily tuning the nuclear SLD and the applied magnetic field offering, in principle, additional control of the magnetic contribution. 

\end{abstract}
 
\begin{highlights}

\item Nuclear and magnetic SLD of amorphous Fe$_x$Co$_y$Al$_8$Zr$_2$ can be tuned largely independently via composition.

\item Provides a practical route for designing magnetic reference layers with targeted nuclear and magnetic SLDs.

\item Matching bilayer SLDs and thicknesses suppresses Bragg peaks, isolating interfacial scattering and revealing a $\sim1$~nm interfacial magnetic dead layer.

\end{highlights}
 
\begin{keywords}
Polarised Neutron Reflectometry \sep Thin Film \sep Sputter Deposition
\end{keywords}
 
\maketitle
 
\section{Background}
 
Neutron reflectometry is a measurement technique used to probe the structure and composition across an interface as a function of depth~\citep{Penfold1990-uv}. Its non-destructive nature and sensitivity to light elements and magnetic structures have made it widely used across fields ranging from soft matter and biology~\citep{Skoda2019-vr,Braun2017-we} to magnetism~\citep{PNRmagnetism} and materials science~\citep{Causer2023-ai}. However, when performing scattering experiments, small contrast can render certain components of a sample effectively invisible. One approach to overcome this is the use of reference layers, as discussed elsewhere~\citep{De_Haan1995-hw,Dikaia2024-gt,Ankner1999-rn,Majkrzak1995-ad,Majkrzak2011-ww,Kirby2012-bu}. The two key parameters that determine the utility of a reference layer are its thickness and scattering length density (SLD). The SLD of a material comprises a nuclear contribution (nSLD), determined by its atomic composition and density, and, in magnetic materials, an additional spin-state-dependent magnetic contribution (mSLD), as defined in Section~\ref{sec:pnr} (Eq.~\ref{eq:sld}). The mSLD adds to the nSLD for one neutron spin state and subtracts from it for the other. By measuring a sample containing a magnetic reference layer (MRL) using polarised neutron reflectometry (PNR), two different contrasts of the same sample can be probed, potentially enhancing the information that can be gained from a given sample~\citep{De_Haan1995-hw}.
 
The nuclear scattering length of an element varies non-monotonically with atomic number (and isotope), and is uncorrelated to its magnetic scattering length, meaning that the nSLD and mSLD of a material can be viewed as independent from one another. Within practical physical and chemical constraints, this independence should allow a wide range of nSLD--mSLD combinations to be realised by combining different elements, offering considerable flexibility in the design of reference layers. While it is difficult to define the full range of all achievable SLDs across all possible materials, and elemental and isotopic compositions, the values observed in single-element materials provide useful context here. For example, the nSLDs of Ti and Fe are $-1.95 \times 10^{-6}$ and $8.0 \times 10^{-6}$~\AA$^{-2}$, respectively, while mSLDs range from $0$~\AA$^{-2}$ in non-magnetic materials to approximately $5 \times 10^{-6}$~\AA$^{-2}$ in Fe.
 
MRLs can be grown as polycrystalline, single-crystalline, or amorphous, with the resulting structure depending on the elemental composition and the deposition conditions, both of which have implications for the utility of a reference layer. Polycrystalline layers can be deposited across a wide range of compositions and substrates, but the presence of grain boundaries and crystallographic disorder can introduce thickness variation, structural roughness and inhomogeneous magnetic behaviour, complicating the interpretation of reflectivity data. Single-crystalline layers offer a high degree of structural and magnetic homogeneity, often with well-defined magnetisation axes that are particularly advantageous in polarised neutron reflectometry; however, producing single-crystal layers requires epitaxial growth on a chosen substrate, imposing significant constraints. In contrast, amorphous layers can be grown on any flat substrate, forming well-defined layers, making them well-suited for reflectometry. However, amorphous alloys typically exhibit lower magnetisation as compared to the magnetic parent material.
 
The compositional freedom afforded by amorphous alloys makes them appealing for MRL design, since the proportion of constituent elements may be varied to target a specific nSLD--mSLD combination. Here, we use amorphous $\mathrm{Fe}_x\mathrm{Co}_y\mathrm{Al}_8\mathrm{Zr}_2$ thin films to demonstrate this tunability directly, varying the Fe:Co ratio while holding the Al and Zr content fixed to preserve an amorphous structure. By characterising a series of these films with polarised neutron reflectometry, we map the accessible nSLD--mSLD space for this combination of elements.
 
\section{Methodology}
 
\subsection{Sample Production}
\label{sec:sample_production}

    \begin{wrapfigure}{r}{0.45\textwidth}
  \begin{center}
    \includegraphics[width=0.42\textwidth]{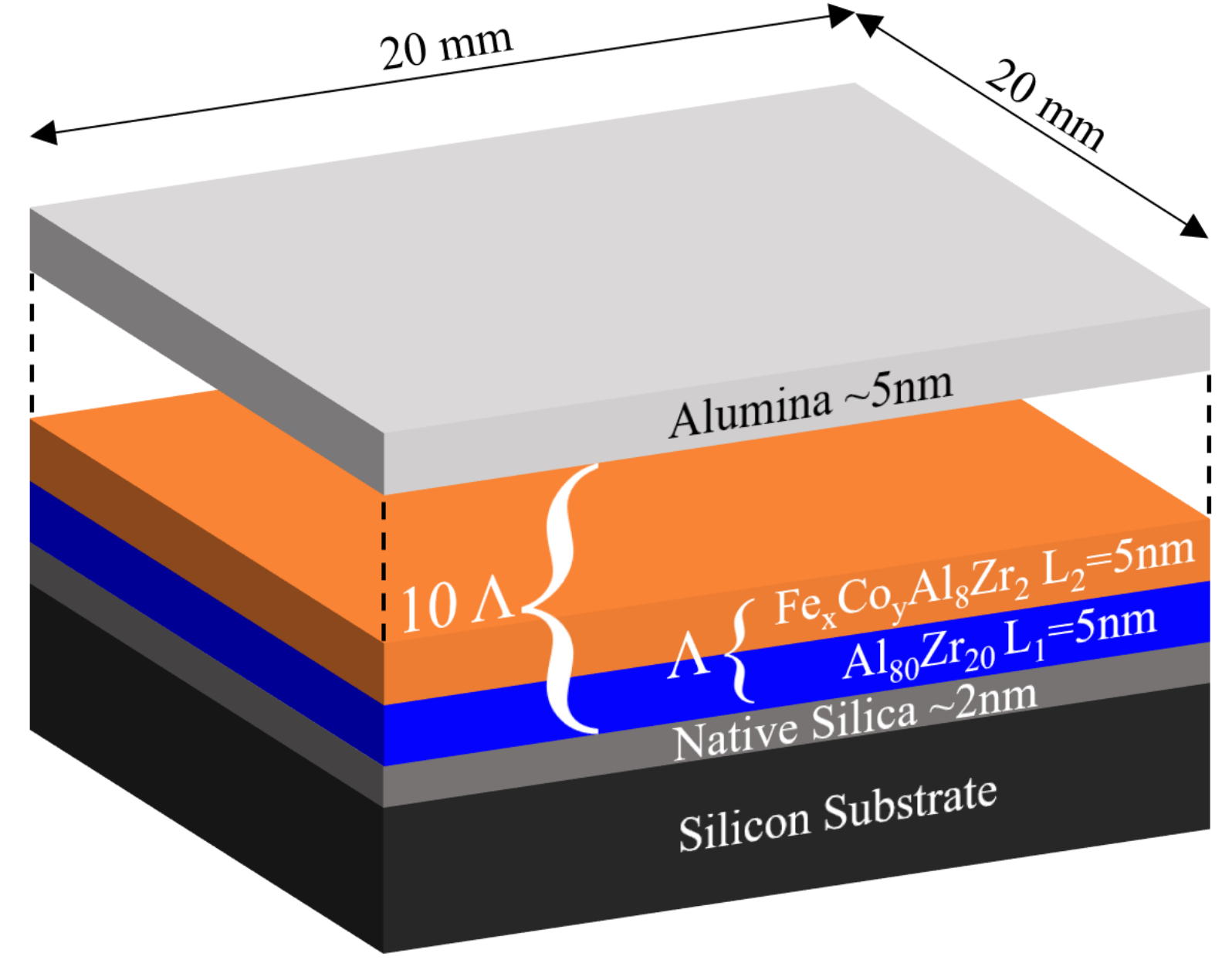}
  \end{center}
 \caption{Intended sample structure following sputter deposition. The bilayer repeat distance, $\Lambda=L_1+L_2$, comprises an Al$_{80}$Zr$_{20}$ spacer layer ($L_1\approx5$~nm) and an Fe$_x$Co$_y$Al$_8$Zr$_2$ magnetic layer ($L_2\approx5$~nm). Ten repeats are deposited on the substrate native oxide and capped with a $\sim5$~nm Al$_2$O$_3$ layer to limit oxidation and environmental degradation of the magnetic layers. The substrate has lateral dimensions of 20$\times$20~mm$^2$.}
 \label{fig:stack}
\end{wrapfigure}

The samples were fabricated in the Materials Physics Laboratory at Uppsala University using DC/RF magnetron sputtering. The growth was carried out under ultra-high-vacuum conditions (base pressure of $\sim10^{-10}$~mbar). Sputtering was performed in high-purity Ar at a pressure of $\sim10^{-2}$~mbar. The Ar (99.999~\%) was further purified using a getter filter. Polished Si(100) wafers, with dimensions of $20\times20\times0.4$~mm$^{3}$ and a native SiO$_2$ layer of approximately 2~nm, purchased from MaTeck GmbH, were used as substrates. Prior to each deposition, the Si substrates were heated to approximately $350~^{\circ}$C under vacuum, followed by Ar flushing before film growth. The target-to-substrate distance was approximately 30~cm. The substrates were mounted off-axis on a rotating holder to improve lateral thickness and compositional uniformity across the deposited films.

The targets (Fe, Co, Al$_{80}$Zr$_{20}$, and Al$_2$O$_3$, 99.95\%) were all 2~inches in diameter and approximately 3~mm thick, with the exception of the Al$_2$O$_3$ target which was 1.5~inches. All the fabricated samples consisted of a silicon substrate, native oxide layer, ten repetitions of an Al$_{80}$Zr$_{20}$ spacer layer ($L_1\approx5$~nm) and an Fe$_x$Co$_y$Al$_8$Zr$_2$ ($L_2\approx5$~nm) magnetic layer, ending with a $\sim5$~nm Al$_2$O$_3$ capping layer in order to limit oxidation and environmental degradation of the magnetic layers. The Al$_{80}$Zr$_{20}$ spacer and Fe$_x$Co$_y$Al$_8$Zr$_2$ magnetic layer form a bilayer of repeat distance $\Lambda=L_1+L_2$. The Fe$_x$Co$_y$Al$_8$Zr$_2$ magnetic layers were prepared by co-sputtering from Fe, Co, and Al$_{80}$Zr$_{20}$ targets, and the Al$_{80}$Zr$_{20}$ spacer layers and alumina capping were prepared by sputter depositing only their respective targets. For the deposition of the magnetic layers, the relative deposition rates were adjusted to vary the Fe:Co ratio between samples while maintaining the relative Al and Zr content. Four different Fe$_x$Co$_y$Al$_8$Zr$_2$ compositions were fabricated, while the overall intended sample structure was kept the same.

Ten repetitions of $\Lambda$ were chosen to enhance the sensitivity of the scattering experiments to the interfacial properties, where the repetition of the layers gives rise to Bragg peaks, enhancing information on the SLD profiles of the samples. By the choice of the layer thicknesses, it is even possible to selectively enhance information from parts of the SLD profile~\citep{Leiner2002-ld,Ahlberg2010-ex}. For example, by choosing $L_1 = L_2$, the second-order Bragg peak ($Q_2$) that exists as a result of $\Lambda$ is minimised.
The Al$_{80}$Zr$_{20}$ spacer layers serve several purposes in the multilayer design: they support the formation of the amorphous magnetic alloy, maintain physical separation between the magnetic layers, and suppress direct magnetic coupling between these. The magnetic layers can therefore be treated as being magnetically independent of each other~\citep{Korelis2012-oj}.

\subsection{XRR}
 
X-ray reflectivity measurements were performed over an angular range of $0.1$--$4^\circ$ using a wavelength of $1.54$~\AA\ corresponding to a momentum-transfer range of $Q = 0.01$--$0.57~$\AA$^{-1}$. A beam size of $1 \times 10$~mm, determined by the slit settings, was used and a footprint correction was applied during the normalisation. The incident beam was attenuated by a factor of 10 for measurements between $0.1$--$2^\circ$, and the measured intensities over this range were scaled by the same factor. The XRR data were fitted in the GenX package~\citep{Glavic2022-oo}, and used to verify that the intended structure had been fabricated, prior to the PNR measurements.
 
\subsection{PNR}
\label{sec:pnr}
Polarised neutron reflectometry (PNR) measurements were performed on the monochromatic Super-ADAM reflectometer \citep{Vorobiev03072015} at the Institut Laue--Langevin (ILL). Measurements were carried out using a neutron wavelength of $5.18~$\AA\ over an incident-angle range of $\theta = 0.1$--$10^{\circ}$, corresponding to a momentum-transfer range of $Q = 0.005$--$0.42~$\AA$^{-1}$. The sample-to-detector distance was 2.5~m. The samples were measured at room temperature under an applied in-plane magnetic field of $500$~mT. The field was sufficiently high to magnetically saturate the FeCoAlZr layers.
 
The polarised neutron reflectometry measurements presented here were analysed using the neutron reflectometry fitting package Refnx~\citep{Nelson2019}. Refnx employs the Abel\`es matrix formalism~\citep{Heavens1960} to calculate a reflectivity profile, the reflectivity as a function of momentum transfer, from a model SLD profile, the SLD as a function of depth. During the fitting procedure, the parameters defining the model SLD profile are iteratively optimised to minimise the difference between the calculated and experimental reflectivity profiles.
 
The SLD at a given depth is given by
\begin{equation}
\mathrm{SLD}_{\pm} = \sum_i n_i b_i \; \pm \; p\,M,
\label{eq:sld}
\end{equation}
\noindent where $n_i$ and $b_i$ are the atomic number density and nuclear scattering length of element $i$, $M$ is the magnetisation of the layer, and $p$ is the magnetic scattering length per unit magnetisation. The sum $\sum_i n_i b_i$ gives the nSLD, while the second term gives the mSLD. The $+$ and $-$ signs in Eq.~\ref{eq:sld} correspond to the magnetic moment of the neutron and the layer being parallel or antiparallel. In the experiments, the external field defining the magnetisation of the sample was kept constant, while neutrons with different spin states were used (spin-up and spin-down).
 
The reflectivity data were modelled using a layered `slab' structure consisting of a bulk silicon substrate, a native silicon oxide layer, a repeated multilayer stack, a surface magnetic layer, an alumina capping layer, and air. The repeated stack comprised nine repeats of an AlZr spacer layer and an MRL separated by interface regions. These interface layers were included to account for the magnetic profile not precisely following the structural profile, while maintaining the same nuclear scattering length density as the repeated-stack MRL. The top MRL was modelled independently from the repeated stack owing to it being capped directly by the alumina layer, rather than another AlZr spacer. The same internal interface layer was also used below the top MRL; a separate interface layer was included between the top MRL and the alumina cap, with the nuclear SLD of the top MRL. A common interfacial roughness was fitted for the oxide, multilayer and alumina interfaces, while the Si/SiO$_2$ roughness was fixed at $0.5$~\AA. The SiO$_2$ thickness and nSLD were fixed at $20$~\AA\ and $3.6\times10^{-6}$~\AA$^{-2}$, respectively. A summary of the parameters is provided in the SI.
 
Simultaneous fitting of the spin-up and spin-down polarized neutron reflectivity datasets was performed using a global objective function, with the nuclear structural parameters shared between spin channels and the magnetic scattering length densities changing sign according to the neutron spin state. The scattering length density models were optimised using dynamic nested sampling~\citep{Skilling2006,Higson2018} implemented in the Dynesty package~\citep{dynesty} to obtain the best fit to the measured reflectometry data. The fitted reflectivity curves and SLD profiles were evaluated at the maximum-likelihood sample. The reported values correspond to the posterior median of the sampled distributions, while the associated uncertainties are given by the 16th and 84th percentiles, representing the lower and upper bounds of the 68\% credible interval, respectively.
 
\section{Results and Discussion}

\subsection{Design of the SLD profile}

A multilayer is characterised by a repeat distance $\Lambda=L_1+L_2$ where $L_1$ and $L_2$ are the thicknesses of the constituent layers. Here we have chosen a sample design where $L_1$/$L_2$ is close to unity, which implies the absence of the second-order structural peak. As explained below, this allows detailed mapping of the changes in the moment of the constituents at the interfaces, providing essential input into the changes in the overall scattering length density profile. This approach has previously been used to obtain information on the hydrogen concentration at interfaces~\citep{Leiner2002-ld}, as well as the effects of boundaries on magnetisation in superlattices~\citep{Ahlberg2010-ex}.
 
\begin{figure}
\centering
\includegraphics[width=\linewidth]{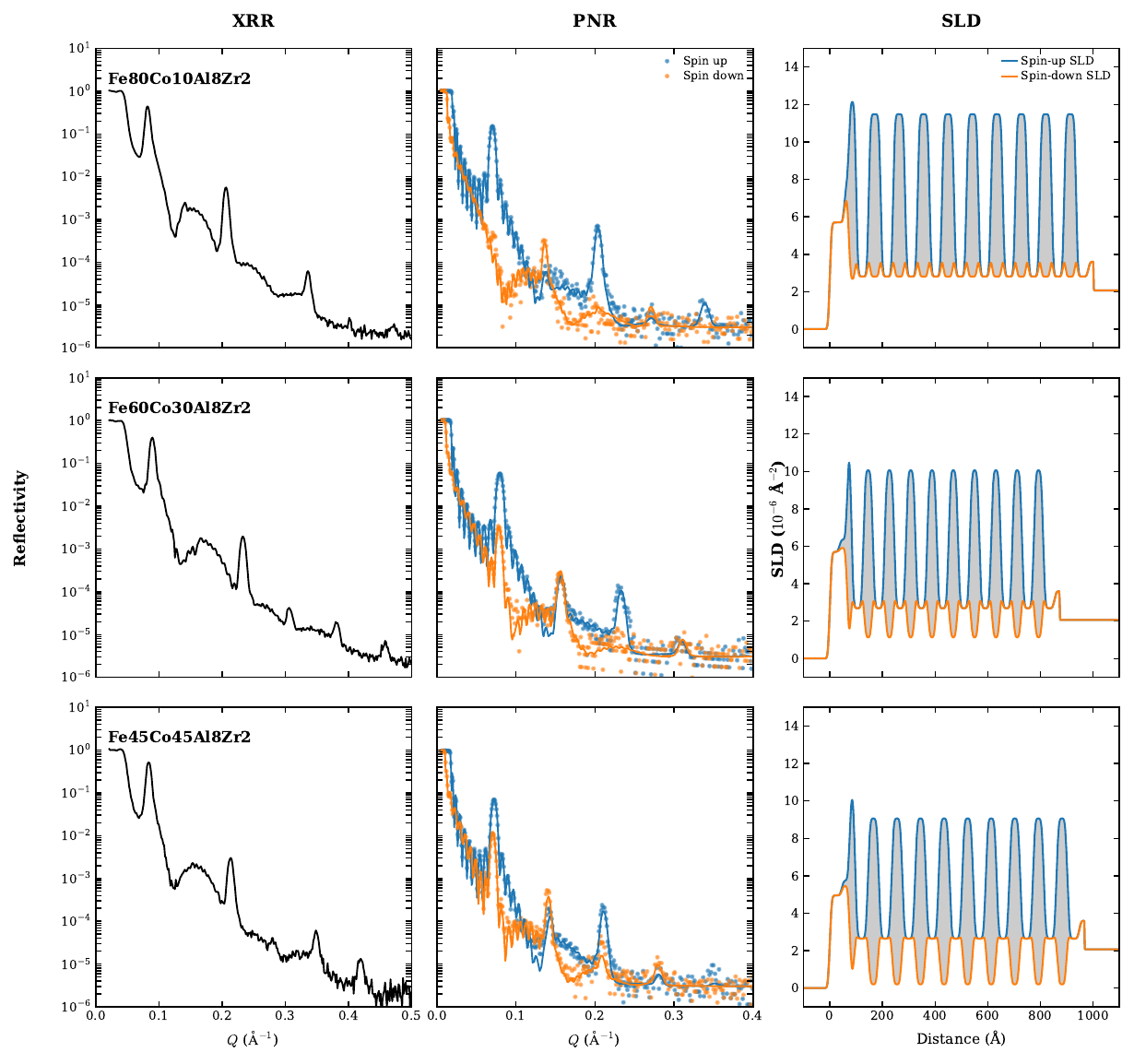}
\caption{Measured XRR data (left), PNR data with calculated reflectivity curves (centre), and fitted spin-dependent SLD profiles (right) for three FeCoAlZr compositions. For Fe$_{80}$Co$_{10}$Al$_8$Zr$_2$, near-perfect matching of the magnetic-layer and spacer-layer SLDs in the spin-down channel suppresses the first-order Bragg peak.}
\label{fig:stack2}
\end{figure}

The measured XRR and PNR data, together with the corresponding best-fit SLD profiles, are shown for three of the samples in Figure~\ref{fig:stack2}, and the fitted parameters are provided in the SI. The repeated bilayer stack present in each sample produces well-defined Bragg peaks in both the XRR and PNR data. For the three samples shown in Figure~\ref{fig:stack2}, the lack of a $Q_2$ peak in the XRR data indicates the successful matching of $L_1$ and $L_2$, although the presence of the fourth-order ($Q_4\approx0.3$~\AA$^{-1}$) and sixth-order ($Q_6\approx0.4$~\AA$^{-1}$) peaks confirms that the layer thicknesses are not identical. For the Fe$_{70}$Co$_{20}$Al$_8$Zr$_2$ sample, $L_1\neq L_2$, and this sample is discussed in the SI. The contribution of the topmost, alumina-capped layer is most easily observed by the wide local maxima at $Q\approx0.18$~\AA$^{-1}$.
 
Figure~\ref{fig:stack2} demonstrates the benefits of the design of SLD profiles. In the Fe$_{80}$Co$_{10}$Al$_8$Zr$_2$ sample, the fitted spin-down SLD of the magnetic layers is $2.80\times10^{-6}$~\AA$^{-2}$, matching the fitted AlZr spacer SLD of $2.81\times10^{-6}$~\AA$^{-2}$ to within $0.01\times10^{-6}$~\AA$^{-2}$, well inside the fit uncertainties. Consequently, the first-order Bragg peak ($Q_1\approx0.08$~\AA$^{-1}$) is absent from the spin-down reflectivity profile. As the Co content increases, the nSLD decreases and the mSLD stays relatively constant, increasing the difference in SLD between the magnetic and spacer layer to $1.57$ and $2.45\times10^{-6}$~\AA$^{-2}$ for Fe$_{60}$Co$_{30}$Al$_8$Zr$_2$ and Fe$_{45}$Co$_{45}$Al$_8$Zr$_2$ respectively, and the intensity of $Q_1$ in the spin-down channel increases correspondingly.
 
The Bragg peaks in the PNR data can also be interpreted directly to gain information about the interfaces between the spacer and magnetic layers that would not otherwise be accessible from the nuclear contrast alone. In the Fe$_{80}$Co$_{10}$Al$_8$Zr$_2$ sample, the absence of $Q_1$ in the spin-down channel reflects the SLD matching described above, which reduces the entire superlattice to a single, SLD-homogeneous component for spin-down neutrons. However, the Bragg peak at $Q\approx0.14$~\AA$^{-1}$ in this channel can be attributed to the interfacial regions at the top and bottom of each magnetic layer, where the reduced magnetic moment leaves a region with a small SLD contrast against the otherwise-matched spacer. Thus, this feature should be understood as the first-order contribution from a length scale corresponding to one half of the repeat distance ($\Lambda$/2). These interfacial regions are present in the SLD profiles of all samples, though most clearly resolved in Fe$_{80}$Co$_{10}$Al$_8$Zr$_2$, when the contrast between the layers is negligible (i.e., no first-order Bragg peak). These regions were necessary to adequately fit the PNR data. The thickness and degree of magnetic suppression in these interfacial regions cannot be uniquely determined, but their integrated magnetic SLD deficit is more strongly constrained by the reflectivity data. Interpreting the fitted SLD profiles as a whole therefore indicates that the regions of reduced magnetic moment are approximately 1~nm in extent, consistent with similar phenomena observed elsewhere \citep{Huang1993-jo,Korelis2012-oj,Kurichenko2026-xp}.
 
\subsection{Tuneability of the nuclear and magnetic SLD}
\label{sec:tuneability}
 
\begin{figure}
\centering
\includegraphics[width=\linewidth]{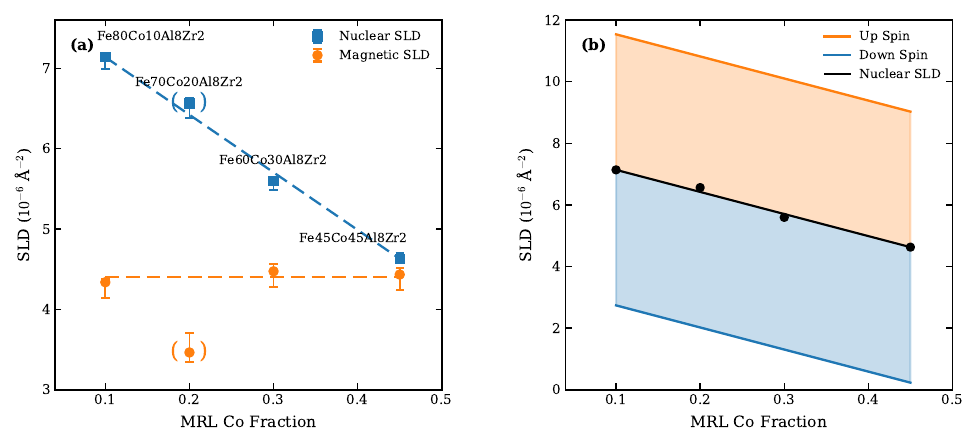}
\caption{(a) Fitted nuclear and magnetic SLDs for the Fe$_x$Co$_y$Al$_8$Zr$_2$ layers as a function of cobalt atomic fraction ($y/100$). Error bars indicate 68\% credible intervals. The dashed lines show an approximation comprising a linear decrease in nuclear SLD and a constant magnetic SLD of $4.4 \times 10^{-6}$~\AA$^{-2}$. The points in parentheses correspond to the Fe$_{70}$Co$_{20}$Al$_8$Zr$_2$ sample, whose layer-thickness ratio differs from the other samples (see SI). \textbf{(b)} Approximate spin-dependent SLD ranges inferred from the linear-nSLD and constant-mSLD approximations in (a), over the investigated range of cobalt fractions. The shaded regions are model-based estimates rather than continuously measured ranges.}
\label{fig:mrlslds}
\end{figure}

Figure~\ref{fig:mrlslds} summarises the refined nuclear and magnetic SLDs of the FeCoAlZr layers as a function of Co fraction across the four samples. The nuclear SLD decreases systematically with composition, from $7.14\times10^{-6}$~\AA$^{-2}$ in Fe$_{80}$Co$_{10}$Al$_8$Zr$_2$ to $4.63\times10^{-6}$~\AA$^{-2}$ in Fe$_{45}$Co$_{45}$Al$_8$Zr$_2$, consistent with the substitution of Fe by Co lowering the average nuclear scattering length of the alloy. By contrast, the magnetic SLD remains relatively constant across the series at approximately $4.3$--$4.5\times10^{-6}$~\AA$^{-2}$, with the exception of Fe$_{70}$Co$_{20}$Al$_8$Zr$_2$, for which a lower value of $3.46\times10^{-6}$~\AA$^{-2}$ was obtained (discussed in the SI). For practical purposes, a useful first-order approximation is to take a constant mSLD of approximately $4.4\times10^{-6}$~\AA$^{-2}$ and a linear dependence of nSLD on Co fraction across the compositional range investigated here; the resulting accessible SLD space, together with this approximation, is illustrated in Figure~\ref{fig:mrlslds}(b). The substantial variation in nSLD with only modest changes in mSLD demonstrates that the two contributions can be controlled largely independently. In addition, at fixed composition, the magnetic SLD can in principle be varied through the applied magnetic field when the film is not magnetically saturated, without changing the nuclear SLD. The measurements presented here were instead performed at a fixed, saturating field of 500~mT.

\subsection{Implications for magnetic reference-layer design}
The results presented here show that an MRL can be intentionally engineered as part of the reflectometry experimental design. Varying the MRL composition allows the spin-dependent SLDs of the MRL to be placed at the values ``ideal'' for gaining information about a sample of interest at the surface \citep{Zubayer2025-zs,Yakimenko2026-ty,Durant2022-sk}, rather than at whatever value a fixed elemental material happens to provide, iron being a common choice for soft matter experiments utilising MRLs \citep{Luchini2025-lf}. For soft matter experiments, it is vital to have a capping layer on top of the MRL to maintain comparable surface chemistry to that of, for example, SiO$_2$, which also protects the reference layer from damage by air, water or buffers \citep{Dikaia2024-gt}. In this context, the MRL provides an additional, spin-dependent contrast variation, complementary to the isotopic substitution routinely used in the sample or subphase itself. As demonstrated here, this additional contrast can itself be designed through alloy composition, rather than fixed by material choice alone. Mapping the accessible nSLD--mSLD space of a candidate alloy, as done here for FeCoAlZr, therefore provides a basis for designing a reference layer around a specific target contrast.
 
\section{Conclusions}
 
Here it has been shown that the SLD profile of a sample can be designed to enhance the information obtained from a neutron reflectometry experiment. Amorphous FeCoAlZr alloys are presented as one candidate material for this purpose, forming smooth and laterally homogeneous films. By varying the Fe:Co ratio, the nuclear SLD was tuned from $7.14$ to $4.63\times10^{-6}$~\AA$^{-2}$, while the magnetic SLD was approximately $4.3$--$4.5\times10^{-6}$~\AA$^{-2}$ for three samples, with a lower value of $3.46\times10^{-6}$~\AA$^{-2}$ for Fe$_{70}$Co$_{20}$Al$_8$Zr$_2$. These results demonstrate substantial compositional control of the nuclear SLD while largely preserving the magnetic contribution; additional, independent control of the latter is possible in principle by varying the applied field below saturation. The ability to design the scattering length density of a material has implications for the design and use of magnetic reference layers, which may act as an alternative or complement to isotopic substitution. One application of this approach is demonstrated here, where matching the SLDs of adjacent magnetic and non-magnetic layers for one neutron spin state isolates the scattering contributions from the interfacial regions between them, which would otherwise be difficult to resolve.
 
\section*{Funding}
This research was supported by the Swedish Research Council (VR) under grants 2021-06151 and 2021-00159.
 
\section*{Conflicts of Interest}
There are no conflicts of interest to declare.
 
\section*{Data Availability}
The XRR and PNR data and the Jupyter notebook used to fit the PNR data are available on Zenodo\cite{zenodo}. The raw neutron data are also available through the ILL data portal under proposal CRG-3229\cite{Stendahl2024-nz}.

\printcredits

\bibliographystyle{cas-model2-names}
 
\bibliography{iucr}

\clearpage

\renewcommand{\thesection}{S\arabic{section}}
\renewcommand{\thesubsection}{\thesection.\arabic{subsection}}
\renewcommand{\thetable}{S\arabic{table}}
\renewcommand{\thefigure}{S\arabic{figure}}

\let\WriteBookmarks\relax
\def\floatpagepagefraction{1}
\def\textpagefraction{.001}

\shorttitle{SI: Design of Nuclear and Magnetic SLD Profiles}

\shortauthors{Barnasas et~al.}

\title[mode=title]{Supplementary Information: Design of Nuclear and Magnetic Scattering Length Density Profiles}

\author[1]{Alexandros Barnasas}[orcid=0000-0002-5523-0714]
\credit{Conceptualization, Data Curation, Investigation, Methodology, Resources, Validation, Writing -- Original Draft}

\affiliation[1]{organization=Department of Physics and Astronomy, Uppsala University,
            addressline={Box 516},
            city={Uppsala},
            postcode={SE-75120},
            country={Sweden}}

\author[1]{Edward J. Stuckey}[orcid=0000-0003-1489-4079]
\cormark[1]
\ead{Edward.Stuckey@physics.uu.se}
\credit{Data Curation, Formal analysis, Methodology, Software, Visualization, Writing -- Original Draft}

\author[2]{Alexei A. Vorobiev}[orcid=0000-0003-3778-4948]
\credit{Data Curation, Investigation, Validation, Writing -- Review \& Editing}

\affiliation[2]{organization={Institut Laue Langevin},
            addressline={Avenue des Martyrs 71},
            city={Grenoble},
            postcode={38000},
            country={France}}

\author[3]{Joshaniel F.K. Cooper}[orcid=0000-0002-4635-2381]
\credit{Formal analysis, Writing -- Review \& Editing}

\affiliation[3]{organization={European Spallation Source ERIC},
            city={Lund},
            country={Sweden}}

\author[1]{Sjoerd Stendahl}[orcid=0000-0002-6571-0746]
\credit{Investigation, Writing -- Review \& Editing}

\author[1]{Bj\"orgvin Hj\"orvarsson}[orcid=0000-0003-1803-9467]
\credit{Conceptualization, Funding acquisition, Methodology, Project administration, Resources, Supervision, Writing -- Original Draft}

\cortext[1]{Corresponding author}

\begin{center}
  {\Large\bfseries Supplementary Information\par}
  \vspace{0.35em}
  {\large Design of Nuclear and Magnetic Scattering Length Density Profiles\par}
  \vspace{1em}
  Alexandros Barnasas$^{1}$, Edward J. Stuckey$^{1,*}$,
  Alexei A. Vorobiev$^{2}$, Joshaniel F.K. Cooper$^{3}$,\\
  Sjoerd Stendahl$^{1}$, Bj\"orgvin Hj\"orvarsson$^{1}$\par
  \vspace{0.8em}
  {\small
  $^{1}$Department of Physics and Astronomy, Uppsala University,
  Box 516, SE-75120 Uppsala, Sweden\\
  $^{2}$Institut Laue Langevin, Avenue des Martyrs 71, 38000 Grenoble, France\\
  $^{3}$European Spallation Source ERIC, Lund, Sweden\\[0.3em]
  $^{*}$Corresponding author: \texttt{Edward.Stuckey@physics.uu.se}\par}
\end{center}
\vspace{1em}

\section{Supplementary methods and results}
\subsection{PNR model and results}

\begin{figure}[htbp]
\centering
\includegraphics[width=\linewidth]{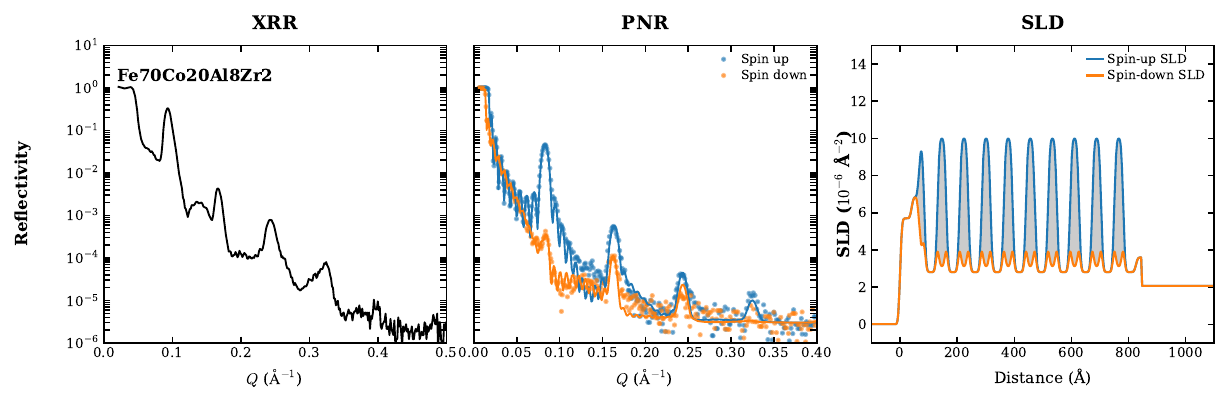}
\caption{XRR and PNR data and fitted scattering length density profile for the Fe$_{70}$Co$_{20}$Al$_8$Zr$_2$ sample. The unequal constituent-layer thicknesses ($L_1\neq L_2$) produce a second-order Bragg peak ($Q_2$) visible in the XRR data.}
\label{fig:mrlFe70Co20}
\end{figure}

The PNR model parameters and fitted values are presented
in Tables~\ref{tab:model_parameters} and~\ref{tab:fit_params},
respectively. The best-fit model, measured reflectivity, and SLD profiles are provided in Figure~\ref{fig:mrlFe70Co20}.

\subsection{Fe$_{70}$Co$_{20}$Al$_8$Zr$_2$}
 
Of the four compositions measured, Fe$_{70}$Co$_{20}$Al$_8$Zr$_2$ shows the largest disparity between the fitted ML and AlZr layer thicknesses, which gives rise to the $Q_2$ peak in the XRR data for this sample, in contrast to the absent $Q_2$ peaks for the other compositions where the two layers are more closely matched in thickness. The fitted magnetic SLD of Fe$_{70}$Co$_{20}$Al$_8$Zr$_2$, $3.46\times10^{-6}$~\AA$^{-2}$, is also notably lower than the $\sim4.3$--$4.5\times10^{-6}$~\AA$^{-2}$ obtained for the other three compositions. The reduced magnetic SLD of the Fe$_{70}$Co$_{20}$Al$_8$Zr$_2$ alloy is attributed to a finite-size effect (e.g., \citealp{Kurichenko2026-xp}) rather than an intrinsic change in the magnetic properties of the alloy owing to the fact that the fitted ML thickness for this sample (27.14~\AA) is smaller than for the other three (32.60--42.72~\AA).

\subsection{X-ray diffraction (XRD)}
\begin{figure}[htbp]
\centering
\includegraphics[width=0.9\linewidth]{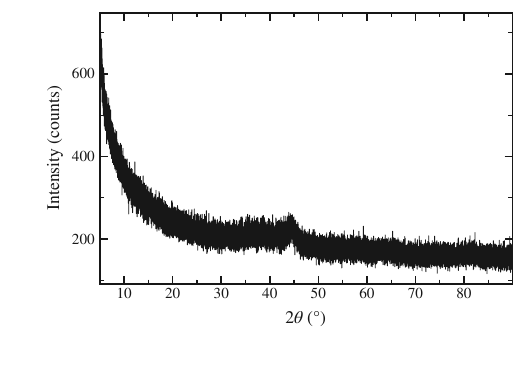}
\caption{XRD of the sample containing magnetic layers of Fe$_{80}$Co$_{10}$Al$_{8}$Zr$_{2}$.}
\label{fig:xrd}
\end{figure}

X-ray diffraction measurements were taken using a Bruker D8 diffractometer, the results of which are shown in Figure~\ref{fig:xrd}. Diffraction patterns were recorded in continuous scanning mode over a range of 2$\theta$ of $5--90^\circ$. The measured diffraction intensities are presented without smoothing or background subtraction.

\begingroup
\footnotesize
\renewcommand{\arraystretch}{1.15}
\setlength{\tabcolsep}{4pt}
\setlength{\LTleft}{0pt}
\setlength{\LTright}{0pt}

\begin{longtable}{@{}
    >{\raggedright\arraybackslash}p{0.28\linewidth}
    >{\raggedright\arraybackslash}p{0.13\linewidth}
    >{\raggedright\arraybackslash}p{0.12\linewidth}
    >{\raggedright\arraybackslash}p{0.38\linewidth}
    @{}}

\caption{Parameters used in the neutron reflectometry models.
Initial values and fitting bounds are provided. Parameters with
specified bounds were varied during optimisation, while parameters
without bounds were held fixed. SLD values are given in units of
$10^{-6}$~\AA$^{-2}$.}
\label{tab:model_parameters}\\

\toprule
\textbf{Parameter} &
\textbf{Initial value} &
\textbf{Bounds} &
\textbf{Description} \\
\midrule
\endfirsthead

\multicolumn{4}{@{}l}{\textit{Table \thetable{} (continued)}}\\
\toprule
\textbf{Parameter} &
\textbf{Initial value} &
\textbf{Bounds} &
\textbf{Description} \\
\midrule
\endhead

\midrule
\multicolumn{4}{r}{\textit{Continued on next page}}\\
\endfoot

\bottomrule
\endlastfoot

\multicolumn{4}{@{}l}{\textit{Instrument settings}}\\
\addlinespace[2pt]
Scale factor & 1 & -- & Reflectivity scale factor \\
Background & $3\times10^{-6}$ & -- & Instrument background \\
Resolution ($dq$) & 2.5 & -- & Instrument resolution smearing \\

\addlinespace[5pt]
\multicolumn{4}{@{}l}{\textit{Bulk materials}}\\
\addlinespace[2pt]
Air nSLD & 0 & -- & Air bulk nuclear SLD \\
Si nSLD & 2.07 & -- & Silicon bulk substrate nuclear SLD \\

\addlinespace[5pt]
\multicolumn{4}{@{}l}{\textit{Substrate oxide}}\\
\addlinespace[2pt]
SiO$_2$ thickness (\AA) & 20 & -- & Native oxide thickness \\
SiO$_2$ nSLD & 3.6 & -- & Native oxide nuclear SLD \\
Si/SiO$_2$ roughness (\AA) & 0.5 & -- &
Fixed substrate/oxide interface roughness \\

\addlinespace[5pt]
\multicolumn{4}{@{}l}{\textit{Multilayer structure}}\\
\addlinespace[2pt]
AlZr thickness (\AA) & 50 & 10--100 & Spacer layer thickness \\
AlZr nSLD & 3.5 & 2--7 & AlZr nuclear SLD \\
Interface roughness (\AA) & 0.5 & 0.5--10 &
Shared roughness excluding Si/SiO$_2$ \\

\addlinespace[3pt]
ML thickness (\AA) & 50 & 10--100 & Repeated ML thickness \\
ML nSLD & 3.5 & 1--8 & Repeated ML nuclear SLD \\
ML mSLD & 3.5 & 0--8 & Repeated ML magnetic SLD \\

\addlinespace[3pt]
Top ML thickness (\AA) & 50 & 10--100 &
Independently refined surface ML thickness \\
Top ML nSLD & 3.5 & 1--8 & Surface ML nuclear SLD \\
Top ML mSLD & 3.5 & 0--8 & Surface ML magnetic SLD \\

\addlinespace[5pt]
\multicolumn{4}{@{}l}{\textit{Magnetically diluted interfaces}}\\
\addlinespace[2pt]
Internal dilute thickness (\AA) & 10 & 0--30 &
Internal ML/AlZr interface thickness \\
Internal magnetic fraction & 0.5 & 0--1 &
Fraction of ML magnetic SLD retained internally \\
Outer dilute thickness (\AA) & 10 & 0--30 &
Alumina/top-ML interface thickness \\
Outer magnetic fraction & 0.5 & 0--1 &
Fraction of ML magnetic SLD retained at surface \\

\addlinespace[5pt]
\multicolumn{4}{@{}l}{\textit{Capping layer}}\\
\addlinespace[2pt]
Al$_2$O$_3$ thickness (\AA) & 35 & 5--100 &
Alumina capping thickness \\
Al$_2$O$_3$ nSLD & 3.5 & 3--5.7 &
Alumina nuclear SLD \\

\end{longtable}
\endgroup


\begingroup
\footnotesize
\renewcommand{\arraystretch}{1.45}
\setlength{\tabcolsep}{3pt}
\setlength{\LTleft}{0pt}
\setlength{\LTright}{0pt}

\begin{longtable}{@{}
    >{\raggedright\arraybackslash}p{0.25\linewidth}
    >{\centering\arraybackslash}p{0.175\linewidth}
    >{\centering\arraybackslash}p{0.175\linewidth}
    >{\centering\arraybackslash}p{0.175\linewidth}
    >{\centering\arraybackslash}p{0.175\linewidth}
    @{}}

\caption{Fitted parameters (posterior medians with 68\% credible
intervals) for each sample. Thicknesses and roughnesses are given
in \AA, and SLDs in units of $10^{-6}$~\AA$^{-2}$.}
\label{tab:fit_params}\\

\toprule
\textbf{Parameter} &
\textbf{Fe80Co10Al8Zr2} &
\textbf{Fe70Co20Al8Zr2} &
\textbf{Fe60Co30Al8Zr2} &
\textbf{Fe45Co45Al8Zr2} \\
\midrule
\endfirsthead

\multicolumn{5}{@{}l}{\textit{Table \thetable{} (continued)}}\\
\toprule
\textbf{Parameter} &
\textbf{Fe80Co10Al8Zr2} &
\textbf{Fe70Co20Al8Zr2} &
\textbf{Fe60Co30Al8Zr2} &
\textbf{Fe45Co45Al8Zr2} \\
\midrule
\endhead

\midrule
\multicolumn{5}{r}{\textit{Continued on next page}}\\
\endfoot

\bottomrule
\endlastfoot

Alumina thickness &
$55.50^{+5.06}_{-5.12}$ &
$40.63^{+4.88}_{-4.99}$ &
$39.72^{+18.59}_{-3.92}$ &
$46.26^{+18.35}_{-1.47}$ \\

Alumina nSLD &
$5.680^{+0.020}_{-0.830}$ &
$5.668^{+0.031}_{-1.042}$ &
$5.675^{+0.023}_{-0.919}$ &
$4.939^{+0.242}_{-0.167}$ \\

Interface roughness &
$5.175^{+0.846}_{-0.210}$ &
$5.147^{+0.963}_{-0.320}$ &
$5.434^{+0.588}_{-0.611}$ &
$5.503^{+0.625}_{-0.635}$ \\

Outer dilute thickness &
$17.01^{+3.93}_{-3.92}$ &
$24.99^{+2.46}_{-11.05}$ &
$23.10^{+6.58}_{-15.30}$ &
$28.09^{+1.82}_{-18.71}$ \\

ML nSLD (outer) &
$7.409^{+0.102}_{-0.735}$ &
$6.839^{+0.174}_{-0.457}$ &
$6.161^{+0.372}_{-0.340}$ &
$5.608^{+0.122}_{-0.371}$ \\

ML mSLD (outer) &
$4.756^{+0.144}_{-1.469}$ &
$2.770^{+0.262}_{-1.690}$ &
$5.164^{+0.446}_{-2.521}$ &
$4.837^{+0.281}_{-1.677}$ \\

Outer magnetic fraction &
$0.090^{+0.324}_{-0.034}$ &
$0.017^{+0.451}_{-0.016}$ &
$0.079^{+0.448}_{-0.043}$ &
$0.049^{+0.437}_{-0.030}$ \\

ML thickness (outer) &
$26.55^{+3.88}_{-4.33}$ &
$16.42^{+19.95}_{-1.58}$ &
$15.08^{+12.84}_{-2.43}$ &
$18.90^{+4.41}_{-3.79}$ \\

Internal dilute thickness &
$2.548^{+5.956}_{-0.285}$ &
$3.634^{+4.739}_{-0.242}$ &
$4.058^{+4.413}_{-0.334}$ &
$5.499^{+4.423}_{-0.652}$ \\

ML nSLD (stack) &
$7.139^{+0.046}_{-0.147}$ &
$6.564^{+0.063}_{-0.187}$ &
$5.597^{+0.044}_{-0.112}$ &
$4.630^{+0.073}_{-0.042}$ \\

ML mSLD (stack) &
$4.338^{+0.037}_{-0.196}$ &
$3.464^{+0.241}_{-0.116}$ &
$4.474^{+0.091}_{-0.196}$ &
$4.434^{+0.081}_{-0.198}$ \\

Internal magnetic fraction &
$0.149^{+0.530}_{-0.138}$ &
$0.072^{+0.497}_{-0.068}$ &
$0.103^{+0.438}_{-0.096}$ &
$0.275^{+0.289}_{-0.088}$ \\

AlZr thickness &
$45.24^{+0.48}_{-1.87}$ &
$43.14^{+1.68}_{-0.45}$ &
$40.34^{+1.11}_{-1.10}$ &
$42.32^{+1.47}_{-1.33}$ \\

AlZr nSLD &
$2.806^{+0.038}_{-0.163}$ &
$2.813^{+0.118}_{-0.031}$ &
$2.696^{+0.098}_{-0.044}$ &
$2.643^{+0.096}_{-0.035}$ \\

ML thickness (stack) &
$42.72^{+0.34}_{-8.85}$ &
$27.14^{+0.38}_{-8.27}$ &
$32.60^{+0.47}_{-6.98}$ &
$36.32^{+1.38}_{-7.21}$ \\

\end{longtable}
\endgroup

\clearpage
 
\end{document}